\documentclass{emulateapj}
\usepackage{apjfonts}
\usepackage{epsfig}
\usepackage{mathrsfs}

\journalinfo{The Astrophysical Journal, 737, 1-11, 2011, August 10}
\slugcomment{Received 2011 February 14; accepted 2011 June 22}
\shorttitle{Star Formation in Self-gravity Accretion Disks in AGNs}
\shortauthors{WANG ET AL.}

\def\civ{C {\sc iv}}

\def\elledd{\ell_{\rm Edd}}
\def\esn{E_{\rm 51}}

\def\mgii{Mg {\sc ii}}

\def\ergs{\ifmmode {\rm ergs~ s^{-1}} \else {\rm ergs~s^{-1}}\ \fi}
\def\kms{\ifmmode {\rm km~ s^{-1}} \else {\rm km~s^{-1}}\ \fi}

\def\mbh{M_{\bullet}}

\def\mgii{\ifmmode Mg {\sc ii} \else Mg {\sc ii}\ \fi}
\def\ciii{\ifmmode C {\sc iii}] \else C {\sc iii}]\ \fi}
\def\heii{\ifmmode He {\sc ii} \else He {\sc ii}\ \fi}
\def\niii{\ifmmode N {\sc iii}] \else N {\sc iii}]\ \fi}
\def\nv{\ifmmode N {\sc v} \else N {\sc v}\ \fi}

\def\siggas{\Sigma_{\rm gas}}
\def\sigsfr{\dot{\Sigma}_*}
\def\sunm{M_{\odot}}
\def\sunmyr{M_{\odot}{\rm yr^{-1}}}

\def\lax{{$\mathrel{\hbox{\rlap{\hbox{\lower4pt\hbox{$\sim$}}}\hbox{$<$}}}$}}
\def\gax{{$\mathrel{\hbox{\rlap{\hbox{\lower4pt\hbox{$\sim$}}}\hbox{$>$}}}$}}

\def\nvciv{N~{\sc v}/C~{\sc iv}}
\def\niiiciii{N~{\sc iii}]/C~{\sc iii}]}
\def\niiioiii{N~{\sc iii}]/O~{\sc iii}]}

\def\ergs{${\rm erg~s^{-1}}$}

\begin{document}

\title{Star formation in self-gravitating disks in active galactic nuclei.
I. Metallicity gradients in broad line regions}

\author{
Jian-Min Wang\altaffilmark{1,2},
Jun-Qiang Ge\altaffilmark{1},
Chen Hu\altaffilmark{1},
Jack A. Baldwin\altaffilmark{3},
Yan-Rong Li\altaffilmark{1},
Gary J. Ferland\altaffilmark{4},
Fei Xiang\altaffilmark{1},
Chang-Shuo Yan\altaffilmark{2}
and Shu Zhang\altaffilmark{1}
}

\altaffiltext{1}
{Key Laboratory for Particle Astrophysics, Institute of High Energy Physics,
Chinese Academy of Sciences, 19B Yuquan Road, Beijing 100049, China; email: wangjm@mail.ihep.ac.cn}

\altaffiltext{2}
{National Astronomical Observatories of China, Chinese Academy of Sciences, 20A Datun Road,
Beijing 100020, China}

\altaffiltext{3}
{Physics and Astronomy Department, 3270 Biomedical Physical Sciences Building, Michigan State
University, East Lansing, MI 48824}

\altaffiltext{4}
{Department of Physics and Astronomy, 177 Chemistry/Physics Building,
University of Kentucky, Lexington, KY 40506}


\begin{abstract}
It has been suggested that the high metallicity generally observed in active galactic nuclei (AGNs)
and quasars originates from ongoing star formation in the self-gravitating part of accretion disks
around the supermassive black holes. We designate this region as the star forming (SF) disk, in which
metals are produced from supernova explosions (SNexp) while at the same time inflows are driven by
SNexp-excited turbulent viscosity to accrete onto the SMBHs. In this paper, an equation of metallicity
governed by SNexp and radial advection
is established to describe the metal distribution and evolution in the SF disk. We find that the
metal abundance is enriched at different rates at different positions in the disk, and that a
metallicity gradient is set up that evolves for steady-state AGNs. Metallicity as an integrated
physical parameter can be used as a probe of the SF disk age during one episode of SMBH
activity. In the SF disk, evaporation of molecular clouds heated by SNexp blast waves unavoidably
forms hot gas. This heating is eventually balanced by the cooling of the hot gas, but we show that
the hot gas will escape from the SF disk before being cooled, and diffuse into the BLRs forming with
a typical rate of $\sim 1\sunmyr$. The diffusion of hot gas from a SF disk depends on ongoing star
formation, leading to the metallicity gradients in BLR observed in AGNs. We discuss this and other
observable consequences of this scenario.
\end{abstract}

\section{Introduction}
It is widely accepted that active galactic nuclei (AGNs) and quasars are powered by accretion
onto supermassive black holes (SMBHs) (Lynden-Bell 1969). The standard accretion disk model
(Shakura \& Sunyaev 1973) or its variants that account for the so-called "big blue bump"
(Shields 1978; Malkan 1983; Laor \& Netzer 1989; Sun \& Malkan 1989; Brunner et al. 1997;
Brocksopp et al. 2006; L\"u 2008) show that a typical accretion rate is a fraction of the
Eddington limit. Over the last four decades, much attention has been paid to the radiation
process and dynamics of the accretion disk, but far less is known about the mechanism driving
such an accretion rate. There is little understanding of how the SMBHs are fed at the $\sim 1$
pc scale, or about the outer boundary condition of the Shakura-Sunyaev accretion disk.

Two important clues can help resolve these wide-open issues. The first is the role of the vertical
self-gravity of the outer part of the accretion disk. Paczynski (1978) pointed out that quasar
accretion disks are so massive
that self-gravity dominates. As a consequence, the disk unavoidably collapses into clumps,
among which collisions provide a natural mechanism for viscosity. This was recently stressed again
by Rafikov (2009). However, Kolykhalov \& Sunyaev (1980) suggested that star formation cannot be avoided
in such a massive disk. Since this pioneering work, many attempts have been made to construct models that
incorporate star formation in the disk as well as take account of its roles in the viscosity and related
issues (Sakimoto \& Coroniti 1981; Lin \& Pringle 1987; Shlosman \& Begelman 1987; Shlosman
et al. 1989; Shlosman \& Begelman 1989; Gammie 2001; Goodman 2003; Collin \& Zahn 1999, 2008;
Thompson et al. 2005; Tan \& Blackmann 2005; Nayakshin \& Sunyaev 2005; Nayakshin et al.
2008; and the recent numerical simulations of Hobbs et al. 2011; Jiang \& Goodman 2011). All of these
studies reach a similar conclusion: star formation in the dense accretion disk
is so efficient that the remaining gas cannot adequately feed the SMBHs unless there is an external
torque to transport angular momentum (e.g. Thompson et al. 2005).

The second major clue is the high metal abundances observed in quasar broad-line regions (BLRs), showing
that there is a fast process of metal enrichment in the nuclear regions of AGNs unless the enrichment
has been done before reaching the nuclear regions (Hamann \& Ferland 1993; see a review of Hamann \&
Ferland 1999; Collin 1999). Generally, the metallicity of host galaxies is significantly lower than that
of the BLR, with even the circumnuclear regions of galaxies having only slightly more than solar metallicity 
(Dors et al. 2008). The BLR metallicity correlates with SMBH mass or AGN luminosity (Hamann \& Ferland
1999; Warner et al. 2003; Matsuoka et al. 2011). Recently, Simon \& Hamann (2010) found that the BLR
metallicity does not correlate with star formation rates of host galaxies as traced by far-infrared
emission. These observational indications provide evidence for intense star formation in a massive disk 
associated with the central engine rather than that the BLR metals are transported from the host galaxies.
However, the problem remains that SMBHs must be fed with a sufficiently high
mass rate to produce the observed luminosity. It has seemed irreconcilable
for AGN to simultaneously have {\it both} high metallicity and large accretion rates.

A solution to this puzzle is to employ turbulence excited by supernovae explosions (SNexp) of massive
stars during star formation as a means to transport angular momentum (Chen et al. 2009; Wang et al.
2010; Hobbs et al. 2011). Recently, Wang et al. (2010) stressed the roles of star formation in metal
production in the inflows developed from the inner edge of the dusty torus, driving not only a
sub-Eddington-limit accretion rate of the SMBHs, but also the formation of a compact nuclear star
cluster after multiple AGN episodes. They find that SNexp have two main effects
in the SF disk: 1) they deliver gas inward, reducing local star formation rates; and 2) the top-heavy
initial mass function quickly feeds processed gas back into the gaseous disk, enhancing the metallicity.
The gas supply to the SMBHs is thus maintained with a rate typical of quasars, while at the same time
the metal-rich properties such as the metallicity-luminosity trend can be quantitatively explained
(Wang et al. 2010). The BLR metallicity measured from the \nv/\civ\, ratio (Hamann \& Ferland 1993)
directly reflects the metallicity in the SF disk as long as the BLR continues to be fed from the SF
disk. Quantitative predictions of the properties of the SF disk and the way in which
it feeds the BLR are urgently needed so that observational tests can be made in the future.

This paper is structured as follows. \S2 gives the model of SF disk in light of Toomre's parameter
and justifications for it. \S3 is devoted to building a metallicity equation for the SF disk that
can predict its metallicity gradient and evolution.
\S4 focuses on the development of thermal diffusion of hot gas from the SF disk. We show that the
diffusion rate can reach as high as a few $\sunmyr$ in quasars. Observational consequences
are then discussed in \S5. Conclusions and some further brief
discussion are given in the last section.

\section{The general structure of the star forming disk}
\subsection{Underlying physical processes in the disk}
We start by summarizing the results published to date about star forming
disks. Most of the release of gravitational energy around an SMBH occurs in a region within about
$ 3-100R_{\rm Sch}$ of the black hole,
where $R_{\rm Sch}=2G\mbh/c^2$ is the Schwarzschild radius, $G$ is the gravity constant, $\mbh$ is the
SMBH mass and $c$ is the light speed. This part of the accretion disk is understood much better than
its outskirts. To minimize the uncertainties due to the opacity etc., we start with the Shakura-Sunyaev
disk for the self-gravity radius given by equation (18) in Laor \& Netzer (1989)
\begin{equation}
R_{\rm SG}\approx 511~\alpha^{2/9}M_8^{-2/9}\dot{m}^{4/9} R_{\rm Sch},
\end{equation}
where $\alpha$ is the viscosity parameter, $\dot{m}=\dot{M}/\dot{M}_{\rm Edd}$,
$\dot{M}_{\rm Edd}=2.22\eta_{0.1}^{-1}M_8\sunmyr$, $M_8=\mbh/10^8\sunm$
and $\eta_{0.1}=\eta/0.1$ is the radiative efficiency, assuming Toomre's parameter $Q=1$. This
is very similar to
$R_{\rm SG}\approx 652~\alpha^{14/27}\mu^{-8/9}\kappa_{\rm R}^{2/9}\ell_{\rm Edd}^{8/27}M_8^{-26/27}R_{\rm Sch}$,
where $\kappa_{\rm R}$ is the Rossland opacity, $\mu$ is the mean molecular weight,
$\ell_{\rm Edd}=L_{\rm Bol}/L_{\rm Edd}$ is the Eddington ratio and $L_{\rm Bol}$ is the bolometric luminosity
(Collin \& Hur\'e 1999). Beyond this radius, the disk is unstable to self-gravity, inducing star
formation. In the self-gravitating regions, the radiation from star formation is given by
$L_*=\eta_*\dot{{\cal R}}c^2$, where $\dot{{\cal R}}$ is the star formation rate and $\eta_*$ is the efficiency
of converting mass into radiation through star formation. In the same regions, the dissipated gravitational
energy bound by the SMBH is given by $L_{\rm acc}\approx R^{-1}R_{\rm Sch}\dot{M}c^2$.
Comparing the two energy sources, we find that stellar energy dominates over the gravitational at
a critical radius $R_{\rm C}\approx \eta_*(\dot{M}/\dot{{\cal R}})R_{\rm Sch}
=10^3\eta_{-3}(\dot{M}/\dot{{\cal R}})R_{\rm Sch}$, where $\eta_{-3}=\eta_*/10^{-3}$ ($\eta_*=10^{-3}$ for
a Salpeter initial mass function in Thompson et al. 2005). This clearly
states that $R_{\rm SG}\sim R_{\rm C}$ for a SF disk with $\dot{{\cal R}}\sim \dot{M}$, implying
that the self-gravitating regions are governed by stellar energy rather
than the gravity of accreting gas bound by SBMHs.

The underlying physics mainly involves several dynamical processes coupling with thermal ones. Fragmentation
into clouds happens when the cooling timescale $t_{\rm cool}\le 3\Omega^{-1}$, where $\Omega$ is the
rotation frequency (Gammie 2001). For a Keplerian rotating disk, $\Omega^{-1}\approx 15.0~r_4^{3/2}$yr, where
$r_4=R/10^4R_{\rm Sch}$, and
$t_{\rm cool}=n_{_{\rm H_2}}kT/\Lambda_{\rm H_2}(n_{_{\rm H_2}},T)\approx 5.5~T_3^{-2.2}$yr, where
$\Lambda(n_{_{\rm H_2}},T)=4.2\times 10^{-31}n_{_{\rm H_2}}T^{3.3}{\rm erg~s^{-1}~cm^{-3}}$ is the cooling
function and $n_{\rm H_2}$ is the density of hydrogen molecules (Smith \& Mac Low 1997). Here estimates of
$t_{\rm cool}$ are conservative since we do not include the metallicity. We find that the fragmentation of
the disk is inevitable, leading to star formation.

The self-gravitational instability drives collisions among clouds, leading to effective viscosity (e.g.
Duschl \& Britsch 2006), but it could be less important than the viscosity due to turbulence produced by
the SNexp. The dispersion velocity obtained from the fragmentation is given by
$V_{\rm disp}=(GM_{\rm cl}\Omega)^{1/3}(t_{\rm coll}/t_{\rm grav})^{1/2}$ (Collin \& Zahn 2008, hereafter CZ08),
where $M_{\rm cl}$ is the cloud mass, and $t_{\rm coll}$ and $t_{\rm grav}$ are the collision
and gravitational interaction between two clouds, respectively. For massive clouds with a mass of $10^3\sunm$,
we have $V_{\rm disp}\approx 156~r_4^{-1/2}\kms$, which is too small compared with the Keplerian rotation
$V_{\rm Kep}=2.1\times 10^3~r_4^{-1/2}\kms$ to excite strong enough turbulence for transportation of angular
momentum in the disk ($<1$ pc). The SNexp-driven turbulent velocity is of the order of the Keplerian velocity
(see \S3.1) and is regarded as the main sources of turbulence, that is to say, $\alpha\approx 1$ is suggested
by 2D numerical simulation (Rozyczka et al. 1995). We note that CZ08 underestimate the SNexp-driven viscosity
since the SNexp would be anisotropic in the fast rotating disk (see detailed arguments in \S3.1).

Detailed processes for forming stars in the disk were studied by Collin \& Zahn (1999, hereafter CZ99) and
CZ08. They considered the following processes: accretion from the SF disk to form the stars, stellar winds
subsequently shutting off such accretion flows, the effects of the accretion process opening gaps in the SF
disk around the forming stars, and migration of stars toward the SMBH.  CZ99 argue that these stars may be
driven to migrate toward the black hole by density waves in the disk, but may be influenced by the cavity opened
by the stellar wind. If the timescale of migration toward the SMBH is shorter than the evolutionary timescale
of main sequence stars, the SNexp-driven turbulence does not work. However, there is actually another mechanism
to stop migration toward the SMBHs, which is that the accretion of stars from the disk increases not only its
mass, but also the angular momentum of the stars. The gain of angular momentum makes the stars migrate outward.
The accretion timescale of stars is $t_{\rm acc}\approx 10^4~(m_*/10\sunm)^{-1}$yr inside the disk environment
(Artymowicz et al. 1993). Comparing $t_{\rm acc}$ with $t_{\rm migr}$ given by equations (28) and (29) in CZ99,
we find that the stars stay at approximately the location where they were born since generally
$t_{\rm acc}<t_{\rm migr}$ works, where $t_{\rm migr}$ is the radial migration timescale. This allows us to
neglect the migration of stars, without obvious influence on the evolution of the stars.

Finally, stars reach their maximum mass when stellar winds balance their accretion from the disk (CZ08)
and they then maintain that mass until they explode as SN. The complicated interplay between these processes
leads to uncertainties in the maximum and minimum
masses of stars. The initial mass function of stars is not obtained in a self-consistent way, but the
uncertainties arising from accretion and stellar winds remain as free parameters in the initial mass function
(see \S3.2).

{
\centering
\figurenum{1}
\includegraphics[angle=-90,scale=0.75]{fig1.ps}
\figcaption{\footnotesize
Structure of the self-gravitating disk, from Eqs (2), (3) and (4). Here we use  $\mbh=10^8\sunm$,
$\kappa=\kappa_{\rm es}= 0.4{\rm cm^2g^{-1}}$, $\alpha=0.1$ and $\elledd=0.3$. The surface density
rate of star formation is very similar to the numerical results given by Wang et al. (2010). For simplicity,
dependence of the opacity on the metallicity is neglected in the calculations.}
\label{disk_density}
\vglue 0.1cm
}

\subsection{The model of a star forming disk}
Wang et al. (2010) computed numerical models of the structures of star forming disks. In order to carry out
an analytical study of the effects of the SF disk, we use a simplified version of its structure,
which are approximately the simple version of a self-gravitating disk (e.g. Goodman 2003). The
smoothed structure of the self-gravitating disk is determined by Toomre's parameter $Q$, and is
justified in this section. It has
been demonstrated by Thompson et al. (2005) that such a disk is supported by the radiation pressure
due to star formation, yielding $Q=1$. The surface density $\Sigma_{\rm gas}$\ and the height of
the SF disk $H_{\rm d}$\ are given by equations (18) and (20) in Goodman (2003)
\begin{equation}
\Sigma_{\rm gas}=1.6\times 10^9~\left(\frac{\elledd}{\alpha_{0.1}\eta_{0.1}}\right)^{1/3}
                 M_8^{-2/3}r_3^{-3/2}~\sunm~{\rm pc^{-2}},
\end{equation}
and
\begin{equation}
H_{\rm d}=4.6\times 10^{-5}
          \left(\frac{\elledd}{\alpha_{0.1}\eta_{0.1}}\right)^{1/3}M_8^{4/3}r_3^{3/2} ~{\rm pc},
\end{equation}
where $r_3=R/10^3R_{\rm Sch}$. The validity of this smooth model of a star forming disk is discussed below.

Fragmentation of the disk depends on the rotation, density, temperature, metallicity and cooling
of individual clouds, whose necessary condition is generally fulfilled as discussed in the previous
section. Details of the star formation process and macrophysics are
poorly understood (McKee \& Ostriker 2007), especially in the accretion disks. Here we will use the
empirical Kennicutt-Schmidt law of star formation rates (Kennicutt 1998). In NGC 5194 (Kennicutt et
al. 2007) and NGC 4252 (Rahman et al. 2010), this law works over all scales from the entire galaxy to
the nuclear region, although in M33 (Onodera et al. 2010) it does appear to break down at the scales
of giant molecular clouds. As a way to move ahead and test the observational consequences of star
formation, we will assume that the Kennicutt-Schmidt law is at least approximately correct for the
accretion disk. The results found here will still be qualitatively correct unless the star formation
rates are completely unrelated to the gas surface density. According to the Kennicutt-Schmidt law,
$\dot{\Sigma}_*=2.5\times 10^{-4}\Sigma_{\rm gas}^{1.4}\sunmyr{\rm kpc^{-2}}$, where $\Sigma_{\rm gas}$
is the surface density of gas in unit of $\sunm~{\rm pc^{-2}}$, and we have the surface density rate
of star formation
\begin{equation}
\sigsfr=1.9\times 10^3~\left(\frac{\elledd}{\alpha_{0.1}\eta_{0.1}}\right)^{0.47}
M_8^{-0.93}r_3^{-2.1}~\sunmyr{\rm pc^{-2}}.
\end{equation}
We find that $\dot{\Sigma}_*$ depends more strongly on the SMBH mass (as $\mbh^{-0.93}$) and the
distance to the SMBHs than on the Eddington ratio (as shown in Figure 1). This leads to the
conclusions that 1) AGN with less massive SMBHs may have different properties from ones with
more massive SMBHs, yielding observational implications  such as
narrow line Seyfert 1 galaxies; and 2) there will be a strong $\sigsfr$ gradient within individual
objects that will lead to a steep metallicity gradient in the SF disk though there is radial advection
as shown in \S3.1, resulting in a difference
in the metallicity measured by different indicators. We discuss these two points in the following
sections.

The clumpy SF region is composed of molecular clouds, evolving stars, SNexp (including the SN shells),
and diffuse hot gas. The structure of the disk is determined by the mutual interaction among these
components (McKee \& Cowie 1975). For simplicity, we use a smooth model of a self-gravitating disk
to represent in analytical form the averaged properties of the disks, but we explicitly consider a
clumpy structure when we determine the way in which gas is heated by SNexp to diffuse into the broad
line regions. The validity of the ``smooth" thin disk can be justified by the fact that the
characteristic distance ($d_{\rm cl}$) among clouds in the star forming regions is much less than the
global size of the disk. The self-gravitating timescale $t_{\rm collapse}=\left(G\rho_c\right)^{-1/2}$,
where $\rho_c$ is the density of clumps, should not be longer than the crossing timescale of turbulence
through a cloud $t_{\rm tur}=l_c/V_{\rm tur}$, where $l_c$ and $V_{\rm tur}$ are the characteristic
length of clouds and turbulence velocity. Otherwise, the cloud will be destroyed to form stars. This
yields the density of clouds as $\rho_c=G^{-1}\left(V_{\rm tur}/l_c\right)^2$. We then have
$d_{\rm cl}=\left(\rho_c/\bar{\rho}\right)^{1/3}l_c=V_{\rm tur}^{2/3}\left(\alpha H/G\bar{\rho}\right)^{1/3}$,
where $\bar{\rho}$ is the mean density of the disk and $l_c\approx \alpha H$ is used. For typical values of
the disk, $\bar{\rho}=10^{-12}{\rm g~cm^{-3}}$, $V_{\rm tur}=10^3\kms$ and $H=10^{15}$cm at $R=10^4R_{\rm Sch}$,
and we have $d_{\rm cl}\approx 2.5\times 10^{16}$cm, which is significantly smaller than the radius of the disk.
The ``smoothed" model is a quite good approximation of the star forming disk\footnote{We note that the SF disk
could be influenced by the stars for the extreme case that the mass of stars is very large.}.

We should point out that a constant opacity is used for the self-gravitating disk as by Goodman (2003). Collin
\& Hur\'e (1999) study the influence on the disk of opacity, as a function of density and temperature. They
show that the surface density of the disk is given by $\Sigma_{\rm gas} \propto \kappa_{\rm R}^{1/7}$ and
$\Sigma_{\rm gas} \propto \kappa_{\rm R}^{1/9}$ for solar abundance and zero-metallicity, respectively (see
equations 8 and 12 in Collin \& Hur\'e 1999). Though the opacity $\kappa_{\rm R}$ can change by a factor of
$10^5$, approximating the opacity as a constant only leads to a factor of a few uncertainty in the surface
density. We note that the disk structure displays several solutions with thermal instability etc. depending
on the opacity (Hur\'e et al. 1994). The resulting star formation in the disk is not very sensitive to the
opacity based on the simplified model of Goodman (2003), but we keep these approximations in mind.

Finally, we note that star formation in the disk is a self-regulating process in light of the Toomre's
parameter, as argued by CZ99. In this paper, we neglect this dependence of star formation, but use the
Kennicutt-Schmidt law for simplicity. Although the Kennicutt-Schmidt law might be modified somehow in
the disk, this approach still illustrates the effects of star formation there. We would like to point out
that star formation takes place in the regions with a Keplerian velocity of  $(1-3)\times 10^3{\rm km~s^{-1}}$.
Detailed process of star formation needs an efficient way to get rid of the angular momentum of the protostars,
perhaps, magnetic fields as a potential mechanism could be invoked to do so.

\section{Metallicity within the star forming disk}
\subsection{SNexp evolution in the SF disk}
In the SF disk, the evolution of a SNexp (here meaning the SN shell) fully depends on its surroundings.
It may conveniently be divided into three stages (e.g. Wheeler et al. 1980 and Shull 1980):
1) rapid ejection followed by expansion with a nearly constant velocity; 2) adiabatic expansion
in the Sedov-phase; 3) the snowplow phase governed by radiation loss when a dense shell forms at
the outer boundary. A SNexp in the SF disk is different from the well-known case for a homogeneous
medium, because the SF disk is much denser so that ``blowouts" in the direction perpendicular to
the disk will be very rare. Rozyczka et al. (1995) studied the evolution of SNexp in a disk within
a 1pc scale through 2-D numerical simulations, and found that the SNexp can in principle provide
turbulent viscosity of $\alpha\sim 0.1$. Here we only give a rough estimate of the SNexp evolution
in the SF disk. According to Shull (1980), for an isotropic SNexp, its Stage 1
ends when the swept mass is comparable with the ejecta of the SNexp, yielding
the radius of the SNexp at this phase $R_{\rm S1}=4.1\times 10^{-5}n_{14}^{-1/3}M_{\rm ej,0}^{1/3}$pc,
where $n_{14}=n/10^{14}{\rm cm^{-3}}$ and $M_{\rm ej,0}=M_{\rm ej}/10^0\sunm$ is the ejected mass. The
duration of Stage 1 is given by $t_{\rm S1}=4.0\times 10^{-3}n_{14}^{-1/3}$yr if the initial ejection
velocity is $10^4$km s$^{-1}$. It is found that $R_{\rm S1}\lesssim H_{\rm d}$ and $t_{\rm S1}\ll t_{\rm K}$,
where $t_{\rm K}=2\pi/\Omega=8.8~M_8^{-1}r_3^{3/2}~{\rm yr}$ is the timescale of the Keplerian
rotation. Stage 1 of a SNexp is not able to be break out of the SF disk (Rozyczka et al. 1995). Wheeler
et al. (1980) demonstrate that the Sedov-phase does not exist provided the density of surrounding gas is
larger than $5\times 10^3\esn^{3/2}\left(M_{\rm ej}/10\sunm\right)^2~{\rm cm^{-3}}$, which is the
case here. Clearly, the SNexp directly enters Stage 3 without radiation loss. Most of the kinetic
energy of the SNexp will be channeled into the kinetic energy of clumps. The maximum radius of the
SNexp is typically $R_{\rm SN}\sim R_{\rm S1}\sim 0.1 H_{\rm d}$, implying that the viscosity driven
by the SNexp will be $\alpha\sim0.1-1$, which is consistent with numerical simulation of
Rozyczka et al. (1995).

\begin{table*}
\begin{center}
\caption{\sc{ A Summary of The Model of Star Forming Disk}}
{\small
\begin{tabular}{lll} \hline\hline
Parameter     & Physical meanings & Typical values (for SMBHs with $10^8\sunm$) \\ \hline
$\elledd$     & Eddington ratio  & $0.3$ \\
$\eta$        & radiative efficiency  & $0.1$ (for intermediately spinning SMBH)\\
$\alpha$      & viscosity parameter   & $0.1$ (from turbulence excited by the SNexp)\\
$M_{\rm disk}$& mass of self-gravitating (or SF) disk   & $\sim \pi \siggas R_{\rm out}^2\sim 5\times 10^{6}\sunm$\\
SFR           & star formation rates        & $\sim 1.0~\sunmyr$\\
SN rate       & supernova rates             & $\sim 3.0\times 10^{-2}$ yr$^{-1}$ for the Salpeter IMF\\
$R_{\rm SG}$  & self-gravitating radius of the SF disk & $\sim 10^3R_{\rm Sch}\approx 10^{-2}$pc\\
$R_{\rm in}$  & inner radius of the SF disk & $=R_{\rm SG}$\\
$R_{\rm out}$ & outer radius of the SF disk & $\sim$ sublimation radius of dusty torus $\sim 1.0$ pc \\ \hline
\label{summary_words}
\end{tabular}}
\end{center}
\end{table*}

On the other hand, the kinetic energy of the SNexp heats the surface of clumps through blast shocks
when the SNexp happens. As we show next, the SNexp ensures that evaporation of molecular clouds via shocks
can drive thermal diffusion of the hot gas above the SF disk into the broad line regions.
Table 1 gives a summary of the present model including the typical parameter values for an SMBH with
$10^8\sunm$, which are useful for future observational tests.

There are debates as to the strength of SNexp-driven viscosity in the star forming disk.
CZ08 argue that the disk only receives the momentum of SNexp corresponding to the velocity vectors
inclined by less than $H/R$. However, the fast rotating disk makes the SNexp anisotropic. Considering
the rotation of the disk, the SNexp energy is channeled into two directions: vertical and $\phi$-direction,
whose ratio is characterized by
$E_{\rm SN}^{\phi}/E_{\rm SN}^{\rm H}\sim \left(V_{\rm rot}+V_{\rm SN}\right)^2/V_{\rm SN}^2=16,$
where $V_{\rm SN}\approx 10^3\kms$ is the typical value of SNexp velocity (see equation 4 in Shull 1980)
and $V_{\rm rot}\approx 3000\kms$ is the Keplerian rotation velocity at $R=10^4R_{\rm Sch}$. We find
$E_{\rm SN}^{\phi}\gg E_{\rm SN}^{\rm H}$, that is to say, the SNexp is anisotropic in the fast
rotating disk. This is the reason why the SNexp shell is
elongated as a ``banana" (Rocyzka et al. 1995). Most of SNexp energy will be exhausted along the
$\phi-$direction, triggering the torque to transfer angular momentum outward. A SNexp in the SF disk
is not able to break down the SF disk, but its evolution makes the metallicity locally homogeneous.
The 2D-treatment in Rozyczka et al. (1995) is a good approximation. The viscosity by SNexp is about
$0.1-1$. However, 3D numerical simulations should be carried out in the future to clarify the importance of the
SNexp-driven viscosity in angular momentum transportation the disk. We use a constant $\alpha=0.1$
in the entire SF disk in equation (4).

\subsection{The metallicity equation}
As a natural consequence of the ongoing star formation in the disk as described above, the metallicity
is enriching and evolving along with the feeding of the SMBHs. As shown by the cooling function in
equation (25) (\S4), metals play a key role in the cooling of the hot gas in the SF disk. In order
to find the diffusion of hot gas, we have to calculate the metallicity gradients in the disk. Detailed
calculations of the gradients involve many processes, including stellar evolution, metal production in
SNexp, and the dynamics of the SF disk. Here we assume that stars stay at the same radius where they
are formed until they produce SNexp in the SF disk (i.e. neglecting migration of stars as discussed in
\S2.1). Since the mixing timescale of metals with its
surroundings fully depends on the SN expansion rate, we find the mixing process can rapidly terminate, in a time
$t_{\rm mix}\sim t_{\rm S1}\ll t_{\rm R}$, where the radial advection timescale is given
by $t_{\rm R}\sim R/V_{\rm R}\sim R/\alpha V_{\rm K}\approx 3\times 10^{2}$yr for $\alpha=0.1$ and
$R=0.1$pc. The metal-rich gas then mixes with its surroundings, and is advected inward by the inflows
while simultaneously being evaporated by the SNexp heating.

A complete calculation of the metallicity evolution would individually follow each of the heavy
elements produced in the SF disk, but here we follow only the overall fraction of metals relative
to the progenitor mass, designated $m_Z$ (Woosley et al. 2002). Following Wang et al. (2010),
the fraction of the SNexp progenitor mass returned to the SF disk, $f_c=\Sigma_{\rm SN}/\Sigma_*$,
can be obtained from the initial mass function
\begin{equation}
f_c=\frac{1}{M_*}\int_{m_c=7\sunm}^{\infty}m_*N_*dm_*,
\end{equation}
where $M_*=\int_{m_{\rm min}}^{\infty}m_*N_*dm_*$. Evidently, the factor $f_{\rm c}$ depends on the
IMF. However, the IMF in the SF disk is very poorly known, as even the IMF in normal galaxies is still
a matter of debate (see the extensive reviews of the IMF by Kroupa 2007 and Dav\'e 2008). The state
of the IMF could be a function of the surroundings of the star forming regions (e.g. Krumholz et al.
2009). However, the stellar mass function in the Galactic center could provide a good hint for the
present case. We approximate the IMF as
\begin{equation}
N_*(m_*)=\left\{\begin{array}{ll}
N_1m_*^{-\beta_1}   & {\rm for}~ m_{\rm min}\le m_*\le m_2,\\
                   &                  \\
N_2m_*^{-\beta_2}   & {\rm for}~ m_*>m_2, \end{array}\right.
\end{equation}
where $\beta_1$ and $\beta_2$ are the indices of the IMF, $m_2$ is the critical mass distinguishing the two
power laws of the IMF, and the two normalizations are connected through $N_2=N_1m_2^{\beta_2-\beta_1}$ for
a continuity of the two power-laws. Top-heavy properties of the IMF are reflected by the two parameters
$\beta_1$ and $m_2$. Usually, $\beta_2=2.35$ which is the Salpeter function. Figure 2 shows several possible
IMFs and the dependence of the factors $f_c$ on the IMF. For a larger lower
limit to the critical mass ($m_c$) of stars, these approximations and assumptions work better. For a
typical IMF, we use $f_c=0.6$.

{
\centering
\figurenum{2}
\begin{figure*}
\center{\includegraphics[angle=-90,scale=0.74]{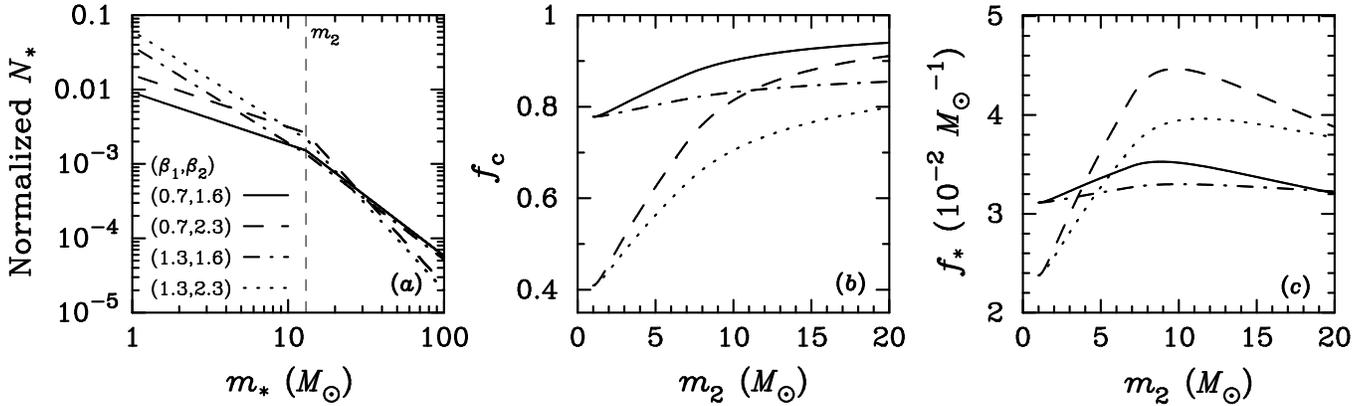}}
\figcaption{\footnotesize ({\em a}) Several different normalized IMFs with $m_2=13\sunm$ based on
Kroupa (2007). ({\em b}) The parameter $f_c$ (equation 5), which is a fraction of the SNexp progenitor
mass returned to the SF disk, for the same IMFs shown in the previous panel. For most cases, more than
$60\%$ of the gas will return to the SF disk to enrich the metallicity. ({\em c}) The parameter $f_*$,
which is the number of SNexp per unit mass of stars formed. We use $f_*=0.02\sunm^{-1}$, a conservative
value, in this paper.
}
\label{fratio}
\end{figure*}
\vglue 0.1cm
}

Since the lifetime of the massive stars is $\sim 10^{6-7}$yr (the minimum mass of stars is $\sim 7\sunm$
for this timescale), much longer than the timescale of the radial motion
($t_{\rm adv}\sim R/V_R\sim R/\alpha V_{\rm K}\sim 10^3$yr for $R=0.1$pc in Wang et al. 2010). Massive
stars produce SNexp in
the accretion inflows, with a delay due to stellar evolution on a time scale of hydrogen burning ($t_*$).
The lag due to evolution effects must be considered for the time-dependent disk.
Taking into account the mass used up in star formation, the net mass decrease due to star formation and
SNexp is given by $\sigsfr-f_c\sigsfr^{\prime}$, where $\sigsfr$ is the surface density of the current star
formation rates at time $t$ and $\sigsfr^{\prime}$ is the rate at an earlier time offset by the lag time $t_*$,
namely, $\sigsfr^{\prime}(t)=\sigsfr(t-t_*)$. The term $f_c\sigsfr^{\prime}$  leads to the metal enrichment
in the SF disk. We should note that metallicity is a cumulative parameter in a region with ongoing star
formation. Considering a ring $\Delta R$ at radius $R$, the gas lost to star formation is
$2\pi R\Delta R(\sigsfr-f_c\sigsfr^{\prime})$,
and the advection of gas due to radial motion is $2\pi R\Delta RV_R\siggas$. We then have
\begin{equation}
\frac{\partial}{\partial t}(2\pi R\siggas)=\frac{\partial}{\partial R} (2\pi RV_R\siggas)
-2\pi R(\sigsfr-f_c\sigsfr^{\prime}),
\end{equation}
yielding
\begin{equation}
\frac{\partial \siggas}{\partial t}=\frac{1}{2\pi R}\frac{\partial \dot{M}}{\partial R}-(\sigsfr-f_c\sigsfr^{\prime}),
\end{equation}
where the inflow rates $\dot{M}=2\pi RV_R\siggas$. We should note that the inflow rate varies with the
radius as a result of the varying amount of gas used up in star formation.
For a metal-enriched gas following equation (8), but with the inclusion of the dissipation of metals by
star formation and production of metals by SNexp, we have
\begin{equation}
\frac{\partial \Sigma_Z}{\partial t}=\frac{1}{2\pi R}\frac{\partial (Z\dot{M})}{\partial R}
-Z\sigsfr + f_cm_Z\sigsfr^{\prime},
\end{equation}
where $Z\dot{M}$ is the metal rate taken by the inflow, and $m_Z=M_Z/M_*$ is a fraction of the metal mass
to its progenitor mass. $m_Z\in [0.1,0.4]$ depends on the progenitors (Woosley et al. 2002). Introducing the
metallicity $Z=\Sigma_Z/\siggas$, we have the metallicity equation
\begin{equation}
\frac{\partial Z}{\partial t}+Z\frac{\partial \ln \siggas}{\partial t}=
\frac{1}{2\pi R\siggas}\frac{\partial (Z\dot{M})}{\partial R}-
\frac{1}{\siggas}(Z\sigsfr-f_cm_Z\sigsfr^{\prime}).
\end{equation}
We would like to stress that this equation can be applied to describe the progression of AGN evolution.
Provided the structure of the SF disk is known, the metallicity and its gradient can be obtained from equation (10)
to diagnose the evolution of accretion rates for both the newly-born and fading AGNs during a single episode.
For an approximately self-similar
evolution of accretion rates during a single episode of AGNs (e.g. Yu et al. 2005), such as in strong starburst
galaxies composite with AGNs (e.g. Laag et al. 2010), this equation can show the time-behavior of the metallicity.
This would be an extremely important key to unveiling the early phase of AGN evolution.

For a stationary SF disk ($\partial \siggas/\partial t=0$) during a single episode of AGNs reaching a steady
luminosity and $\sigsfr^{\prime}=\sigsfr$, we have,
\begin{equation}
\frac{\partial \dot{M}}{\partial R}=2\pi R(1-f_c)\sigsfr,
\end{equation}
and metallicity evolution reduces to
\begin{equation}
\siggas\frac{\partial Z}{\partial t}=\frac{\dot{M}}{2\pi R}\frac{\partial Z}{\partial R}
          +f_c\left(m_Z-Z\right)\sigsfr.
\end{equation}
where the parameters of the SF disk are constant with time, except that the metallicity
is increasing. A simplified version of equation (12) is given by
\begin{equation}
\frac{\partial Z}{\partial t}=V_{\rm R}\frac{\partial Z}{\partial R}
     +f_c(m_Z-Z)\frac{\sigsfr}{\siggas},
\end{equation}
which is imposed by the initial and boundary conditions
\begin{equation}
Z(t=0,R)=Z_0,~~~~(R_{\rm in}\le R\le R_{\rm out});
\end{equation}
and
\begin{equation}
Z(t, R=R_{\rm out})=Z_0,
\end{equation}
where $R_{\rm in}$ and $R_{\rm out}$ are the outer and inner radii of the SF disk, respectively.
Here we assume that the initial condition of metallicity is homogeneous throughout the entire SF
disk.

We would like to point out that for the steady SF disk 1) metallicity as an integrated parameter
in a steady SF disk is evolving, this is one of the most important characteristics of the present
scenario; 2) the metallicity gradient and evolution only depend on the radial transportation if the
star formation law is a linear relation; 3) $Z(t, R)$ clearly relies on the parameters $f_c$ and
$m_Z$, but we neglect their dependence on radius as well as the mass of SNexp progenitors; 4) the
disk structure and metallicity actually are coupled to each other and a self-consistent model should
produce the structure and metallicity simultaneously, although we presume a known structure of the
SF disk in this paper. The largest uncertainties of the present results originate from the star formation
law and the factor $f_c$. We keep these uncertainties in mind, but leave them for a future paper.

\subsection{The formation of a metallicity gradient}

Radial transportation of metals due to advection produces a saturated metallicity gradient
in the SF disk. When the metallicity is saturated (i.e. $\partial Z/\partial t=0$), we have
\begin{equation}
\frac{dZ}{dR}=f_c(Z-m_Z)\frac{2\pi R\sigsfr}{\dot{M}},
\end{equation}
from equations (12) or (13). The analytical solution is given by
\begin{equation}
\ln \left[\frac{m_Z-Z(R)}{m_Z-Z_0}\right]=f_c\int_{R_{\rm out}}^R\frac{2\pi R\sigsfr}{\dot{M}}dR
=f_c\int_{R_{\rm out}}^R\frac{\dot{\Sigma}_*dR}{V_{\rm R}\Sigma_{\rm gas}}.
\end{equation}
As an illustration of the present model for the metallicity gradient, we approximate the numerical
solution (Wang et al. 2010) as $V_{\rm R}=a_1x^{\alpha_1}$, and
$f_c\dot{\Sigma}_*/\Sigma_{\rm gas}=a_2x^{-\alpha_2}$, where $a_1=2.5\times 10^{-7}~{\rm pc~yr^{-1}}$,
$a_2=6.5\times 10^{-8}{\rm yr^{-1}}$, $\alpha_1=0.94$, $\alpha_2=0.54$, $f_c=0.6$, $m_Z=0.4$,
$x=R/R_{\rm out}$ and $R_{\rm out}=1.0$pc (which is the inner radius of the dusty torus).
It follows from equation (17)
\begin{equation}
Z(x)=m_Z-(m_Z-Z_0)\exp\left[\frac{a_2 R_{\rm out}}{\alpha_3a_1}\left(x^{\alpha_3}-1\right)\right],
\end{equation}
where $\alpha_3=1-\alpha_1-\alpha_2$ obtained from the approximation of disk structure. When the
index term is much less than unity, we have the approximation
\begin{equation}
Z(x)\approx Z_0-(m_Z-Z_0)\left[\frac{a_2 R_{\rm out}}{\alpha_3a_1}\left(x^{\alpha_3}-1\right)\right],
\end{equation}
for the outer regions, displaying a power-law gradients as $Z\propto x^{\alpha_3}$. For a boundary
with an extremely low metallicity, the metallicity tends to
$Z\approx m_Z\left[a_2R_{\rm out}\left(1-x^{\alpha_3}\right)/\alpha_3 a_1\right]$.

{
\centering
\figurenum{3}
\begin{figure*}[!]
\center{\includegraphics[angle=0,scale=0.55]{fig3a.ps}
\includegraphics[angle=0,scale=0.55]{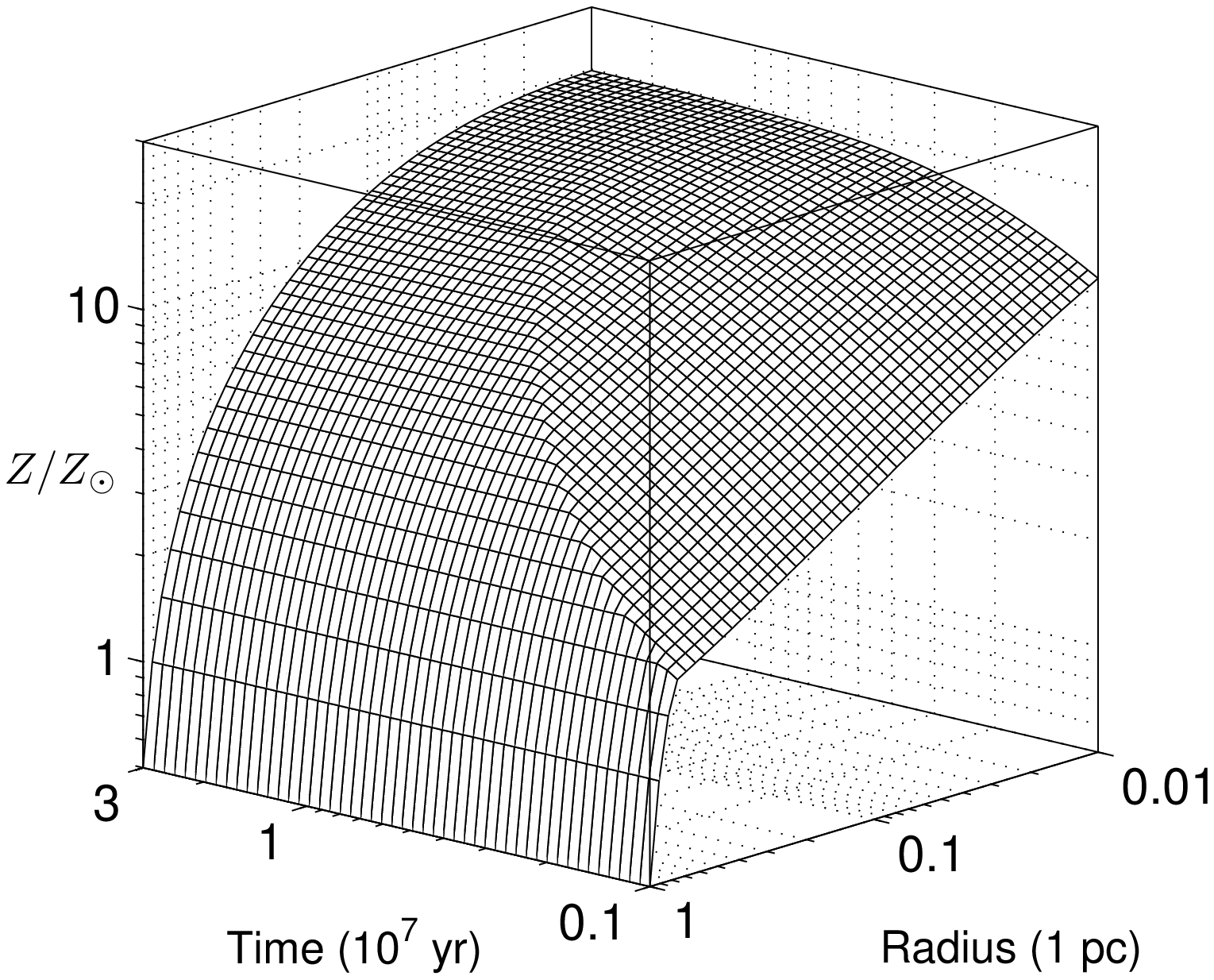}}
\figcaption{\footnotesize {\em Left}
Metallicity gradient of the SF disk when $Z$ is saturated at each radius. The four colored lines are
for different initial metallicity in the SF disk. We find that the gradient of the metallicity is quite
large, so there should be significant difference between the metallicities of the inner and outer parts
of the accretion disk. {\em Right} Metallicity evolution of the SF disk for a given initial metallicity.
The model of metal evolution is only valid after a few times $10^6$yr after the initial star formation.
We use $\mbh=10^8\sunm$ and $\elledd=0.3$.
}
\label{metallicity}
\end{figure*}
\vglue 0.1cm
}

Figure 3 ({\em left}) shows the metallicity and its gradient in the SF disk. The
gradients can be characterized by one power-law part as expressed by the approximation in Eq. (19)
and by a second saturated component. The exact shape of the metallicity gradient depends on the boundary
condition, namely on the metallicity at the inner edge of the torus, but the innermost region of the SF
disk always has the maximum metallicity independent of this boundary condition. The maximum metallicity
is $Z_{\rm max}=m_Z$, which is the metallicity of the SNexp progenitor.

\subsection{Metallicity evolution}
If a SMBH is fed with a constant accretion rate in a single episode, its evolutionary progress
cannot be probed by using the SMBH mass accretion rate, but the metallicity  {\em is} an indicator
of the ongoing star formation. This interesting feature allows us to estimate the {\em age} of the
SMBH activity. The partial differential equation (13) can be analytically solved for the case of a
SF disk with a power-law structure. Employing the approximation of Wang et al. (2010),
$V_{\rm R}=a_1x^{\alpha_1}$, $f_c\dot{\Sigma}_*/\siggas=a_2x^{-\alpha_2}$,
the solution to the partial differential equation (13) is characterized as
\begin{equation}
\frac{x^{1-\alpha_1}}{1-\alpha_1}+\frac{a_1}{R_{\rm out}}t=C_1,
\end{equation}
and
\begin{equation}
\ln (m_Z-Z)-\frac{a_2R_{\rm out}}{a_1\alpha_3}x^{\alpha_3}=C_2,
\end{equation}
where $\alpha_3=1-\alpha_1-\alpha_2$, $C_1$ and $C_2$ are two constants.
We thus have the general solution for the metal evolution as
\begin{equation}
\ln (m_Z-Z)-\frac{a_2R_{\rm out}x^{\alpha_3}}{a_1\alpha_3}={\cal F}\left(\frac{x^{1-\alpha_1}}{1-\alpha_1}
           +\frac{a_1}{R_{\rm out}}t\right),
\end{equation}
where ${\cal F}$ is any continuous function.
Applying the general solution to the initial and outer boundary conditions, we have the solutions
\begin{equation}
\ln\left(\frac{m_Z-Z}{m_Z-Z_0}\right)=\frac{a_2R_{\rm out}}{a_1\alpha_3}
\left\{x^{\alpha_3}-\left[x^{1-\alpha_1}+\frac{a_1}{R_{\rm out}}(1-\alpha_1)t)\right]^{\alpha_3/(1-\alpha_1)}\right\},
\end{equation}
and
\begin{equation}
\ln\left(\frac{m_Z-Z}{m_Z-Z_0}\right)=\frac{a_2R_{\rm out}}{a_1\alpha_3}\left(x^{\alpha_3}-1\right).
\end{equation}

For a given SF disk with typical parameters, Figure 3 {\em right} shows
the evolution of the metallicity gradient. At each radius in the SF disk, the metallicity increases with
time until it reaches a saturation value at a critical time. These saturated points consist of a breaking
curve on the metallicity surface. Radial transportation due to advection balances with the local production
of metals. In the innermost regions of the SF disk, the metallicity reaches its maximum value in a few
times $10^7$yr. It should be noted that the radial transportation is due to the SNexp-turbulent viscosity
(Wang et al. 2010). The metallicity gradient itself is thus a self-organized result without external factors.
The property of metallicity with time is very useful for estimation of star forming disk ages.

Equation 10 describes the
metallicity evolution in the disk, provided that the mass rates of the supplied gas are known. It is
clear that galaxies are undergo repeated episodes of SMBH activity and star formation (e.g. Marconi et al.
2004; Wang et al. 2006; also see the introduction in CZ08). For a steady star forming disk, the equation
reduces to one for metallicity evolution in a single episode of SMBH activity. During that single episode,
the evolutionary phase of SMBH accretion is very hard to estimate from just the accretion rates since they
are roughly constant. However, the property that metallicity monotonically increases
with time allows us to use it as a probe of the age of the AGN outburst. The next section will discuss for
a single episode the diffusion of hot gas driven by SNexp from a steady star forming disk, forming a hot
 ``corona" above the disk.

\section{Thermal diffusion of hot gas from the SF disk}
Interaction between the blast wave of a supernova remnant and interstellar clouds will lead to the
evaporation of clumps, producing hot gas between the clumps (McKee \& Cowie 1975), meaning that
SNexp heating is driving production of hot gas in the SF disk. A stationary evaporation rate from
the SF disk can be reached by the balance between the SNexp heating and the cooling of the hot gas.
Metallicity has a strong influence on the cooling rate of hot plasma (e.g. B\"ohringer \& Hensler
1989). Since metallicity is quite high in AGNs and quasars, we include the effects of metallicity
on the cooling of the hot plasma, and get the diffusion of hot gas self-consistently.

With the inclusion of emission lines, the cooling rate due to thermal emission from hot gas is
approximately given by
\begin{equation}
\Lambda_{\rm cool}^{\rm SF}=2\times 10^{-10}n_6^2T_6^{-1.3}\left(\frac{Z}{Z_\odot}\right)~{\rm
 erg~s^{-1}~cm^{-3}},
\end{equation}
for a range of temperature of $10^5-10^7$K from Fig.1 in B\"ohringer \& Hensler (1989). Here
$T_6=T/10^6$K and $n_6=n_e/10^6~{\rm cm^{-3}}$ are the temperature and density of the
hot gas evaporated in the SF disk.

The heating rate for the same gas depends on the rate of SNexp and the energy injected per SNexp.
Using the IMF given by equation (6), the number of SNexp per unit mass of stars formed is
\begin{equation}
f_*=\frac{1}{M_*}\int_{m_{\rm c}}^{\infty}N_*dm_*.
\end{equation}
Values of $f_*$ depends on the IMF, but it is in a range of $f_*\in [0.02,0.05]\sunm^{-1}$ from Figure
2{\em c}. We use the lower limit value of $f_*=0.02\sunm^{-1}$ in this paper.
The heating rate per unit volume due to SNexp is then given by
\begin{equation}
\begin{array}{ll}
\Lambda_{\rm SN}&=\displaystyle f_*\left(\frac{\dot{\Sigma}_*}{H_{\rm d}}\right)\xi E_{\rm SN}\\
                &   \\
                &=1.0\times 10^{-10}f_{0.02}\xi_{1/6}\dot{\Sigma}_{*,0}H_{14}^{-1}E_{51}
                ~{\rm erg~s^{-1}~cm^{-3}},
\end{array}
\end{equation}
where $f_{0.02}=f_*/0.02\sunm^{-1}$, $\xi_{1/6}=\xi/\frac{1}{6}$ is the fraction of the kinetic energy
converted into thermal energy, $\dot{\Sigma}_{*,0}=\dot{\Sigma}_*/10^0\sunmyr{\rm pc^{-2}}$,
$E_{51}=E_{\rm SN}/10^{51}{\rm erg}$ is the kinetic energy of the SNexp and $H_{14}=H_{\rm d}/10^{14}$cm.
The values of $\xi$ and $f_*$ are both uncertain, and are interrelated. In the following equations they
appear together as the product $f_{\xi}=\xi f_*$. For a given $f_{\xi}$, the higher $f_*$, the smaller
$\xi$. For the minimum $f_{*,\rm min}=0.02\sunm^{-1}$, the parameter is $\xi\approx 1/6$ roughly.
Provided $ f_{\xi}$ is at a level of $\sim 10^{-3}\sunm^{-1}$, which is sufficient for the present model,
the required $\xi$ is only a few percent. Though it is difficult to estimate $f_*$ and $\xi$, the present
model is not sensitive to their exact values.

We can then estimate the temperature of the hot gas from the energy balance
$\Lambda_{\rm cool}^{\rm SF}=\Lambda_{\rm SN}$ in the SF disk. This yields
\begin{equation}
n_6^2T_6^{-1.3}Z_1=0.50~f_{\xi,-3}E_{51}\dot{\Sigma}_{*,0}H_{14}^{-1},
\end{equation}
where $Z_1=Z/Z_{\odot}$ and $f_{\xi,-3}=f_{\xi}/3\times 10^{-3}$. A further constraint on the
condition of the hot gas is that it should not be cooled before escaping from the SF disk. This
requires a critical value of $t_{\rm cool}^{\rm SF}=H_{\rm d}/c_s$, where $c_s=(kT/\mu m_p)^{1/2}$
is the sound speed of the hot gas, and
\begin{equation}
t_{\rm cool}^{\rm SF}=\frac{n_ekT}{(\gamma-1)\mu}\frac{1}{\Lambda_{\rm cool}^{\rm SF}}
               =3.1\times 10^6~\mu_0^{-1}n_6^{-1}T_6^{2.3}Z_1^{-1}~{\rm sec},
\end{equation}
where $k$ is the Boltzmann constant, $m_p$ is the proton mass, $\gamma=4/3$. We then have
\begin{equation}
n_6^{-1}T_6^{2.8}Z_1^{-1}=2.91~\mu_0^{3/2}H_{14}.
\end{equation}
Combining equations (28) and (30), we have the density and temperature of the hot plasma in the SF disk,
\begin{equation}
n_6=0.88~f_{\xi,-3}^{0.65}E_{\rm 51}^{0.65}\mu_0^{0.45}Z_1^{-0.36}\dot{\Sigma}_{*,0}^{0.65}H_{14}^{-0.36},
\end{equation}
and
\begin{equation}
T_6=1.40~f_{\xi,-3}^{0.23}\mu_0^{0.69}E_{51}^{0.23}Z_1^{0.23}\dot{\Sigma}_{*,0}^{0.23}H_{14}^{0.23}.
\end{equation}
Since the hot gas is cooling on a time scale longer than the timescale for escaping from the SF disk,
it forms strong diffusion into regions above the SF disk. The diffusion rates driven by the SNexp are
then given by
\begin{equation}
\begin{array}{ll}
\dot{M}_{\rm diff}&={\displaystyle \int_{R_{\rm in}}^{R_{\rm out}}2\pi c_sm_pn_{\rm h}R dR}\\
        &                          \\
&=\displaystyle 1.8~\int_{x^{\prime}_{_{\rm in}}}^{x^{\prime}_{_{\rm out}}}C_0Z_1^{-0.24}
\dot{\Sigma}_{*,0}^{0.77}H_{14}^{-0.24}x^{\prime} dx^{\prime}~\sunmyr,
\end{array}
\end{equation}
where $C_0=f_{\xi,-3}^{0.77}E_{51}^{0.77}\mu_0^{0.31}$,
$R$ is the radius of the SF disk, $x^{\prime}=R/10^{-0.5}{\rm pc}$, $R_{\rm in}=R_{\rm SG}$
and $R_{\rm out}=1$pc. Equations (31), (32) and (33) are the form solution of the hot
gas in the SF disk in light of $\sigsfr$, $H_{\rm d}$ and the metallicity $Z$.

{
\centering
\figurenum{4}
\begin{figure*}
\center{\includegraphics[angle=-90,scale=0.75]{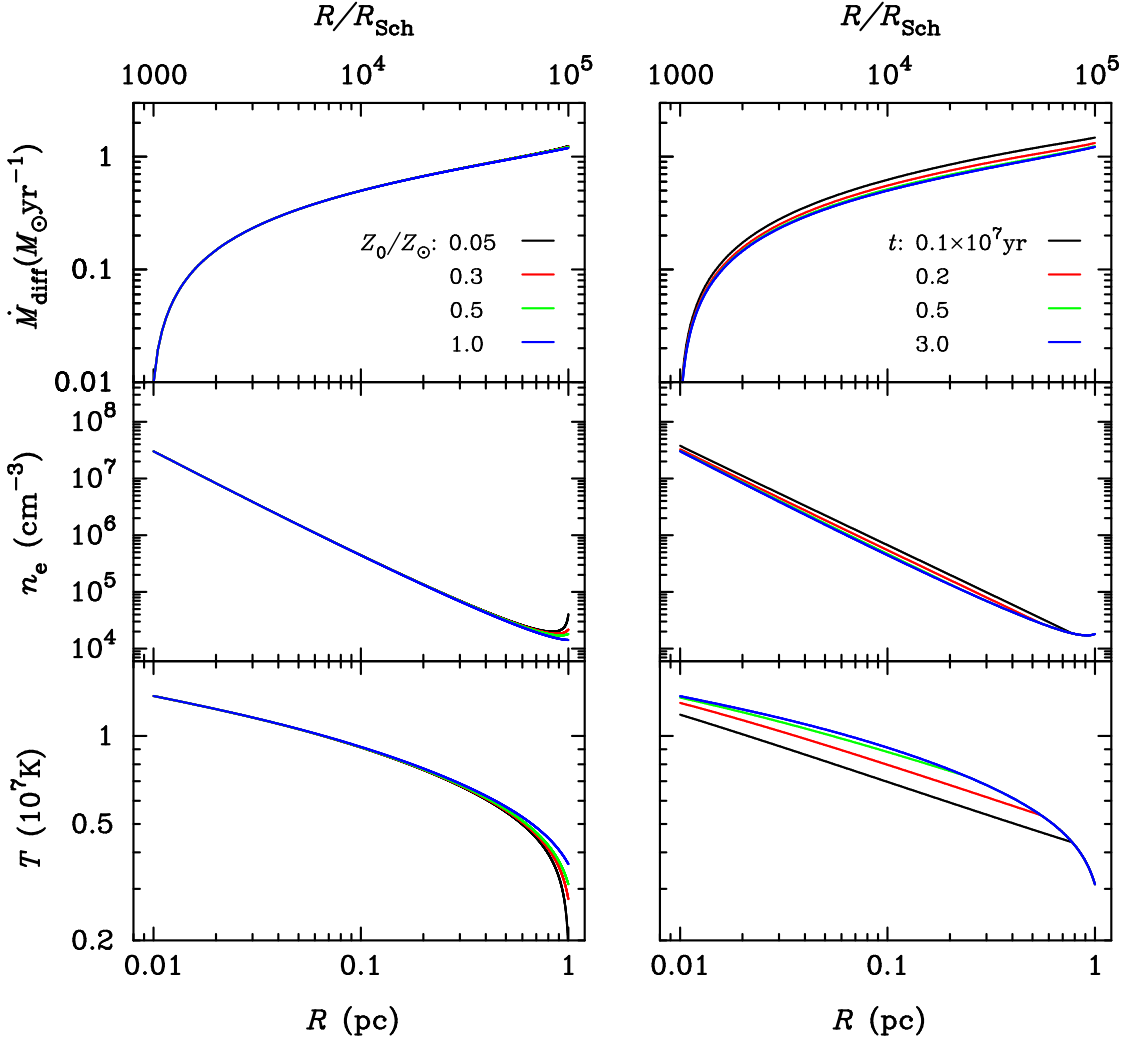}}
\figcaption{\footnotesize
Structures of the hot gas diffused from the SF disk. {\em Left} is for an steady SF disk and
{\em right} for the time-dependent. From bottom to top panels, we show
temperature, density, surface-density of star formation rate and mass injection rate
of the thermal diffusion. We use the Eddington ratio $\elledd=0.3$. The mass injection
rate is from the $[R_{\rm in},R]$ part of the SF disk, the $[R_{\rm in},R_{\rm out}]$
(i.e. Equation 33) is the total mass rates with an order $1\sunmyr$. We assume the initial
metallicity $Z=0.5Z_\odot$ throughout the SF disk (the {\em rgiht} panel).}
\label{outflow}
\end{figure*}
\vglue 0.1cm
}

Inserting equations (3), (4), (18) and (22) into (31-33), we find the structures of the diffused hot
gas for the case of saturated metallicity and the evolution of those structures with increasing
metallicity, respectively. For the saturated metallicity, Figure 4 {\em left} shows the structure of
the diffused hot gas above the SF disk. The {\em left} panel shows that the density and temperature
distribution of the hot gas is insensitive to the boundary metallicity when the metallicity reaches
its saturated state. The temperature of the hot gas in the SF disk is roughly constant with radius,
staying in the range between $(0.5\sim 1.0)\times 10^7$K. We note that the temperature is lower than
the typical Compton temperature of the AGN gas. The density of the hot gas strongly depends on the
radius, falling from roughly $10^7$ to $10^4$cm$^{-3}$ from the inner to the outer part of the disk.
We find that the total mass rates of the hot gas diffusion over the entire SF disk can be as high as
a few $\sunmyr$ if the Eddington ratio is $\elledd=0.3$. We should note that this mass rate could be
the lower limit since we have conservatively taken $f_{\xi}\sim 10^{-3}\sunm^{-1}$. Additionally
the diffusion rates of hot gas are comparable with or exceed the Eddington limit of a $10^8\sunm$
SMBH, but they are still smaller than the rates of the inflows developed from the inner edge of a
dusty torus (see Wang et al. 2010).

Figure 4 {\em right} panel shows the evolution of the hot gas and its diffusion with the metallicity.
We find that temperature and the density of the hot gas vary only slightly with $Z$, but the diffusion
rates vary by a factor of a few. The higher the metallicity, the lower the diffusion rates. For an
individual AGN evolving with metallicity, the thermal diffusion becomes weaker.

It should be pointed out that the diffused hot gas is actually still bounded by the potential of
the SMBH. We can estimate the thickness of the diffused hot gas by assuming vertical static balance,
$H_{\rm diff}\approx c_s/\Omega_{\rm K}=1.4\times 10^{16}~T_6^{1/2}r_4^{3/2}M_8$cm,
$H_{\rm diff}/R\sim 0.05~T_6^{1/2}r_4^{1/2}$ and $H_{\rm diff}\approx 3 H_{\rm d}$, where
$H_{\rm d}$ is the thickness of the SF disk. The
thermal diffused hot gas then meets complicated fates as a result of heating by radiation from the
accretion disk. The hot ionized gas will undergo thermal instability and form cold
clouds in the broad line region since it is continuously supplied through the evaporation driven
by SNexp in the SF disk. The metallicity gradient in the SF disk determines the metallicity in the BLR.
A detailed model of the BLR formed by the hot gas will be the subject of a forthcoming paper (Wang et
al. 2011 in preparation).

Combining equations (3), (4) and (33), we find that the diffusion rates of hot gas
$\dot{M}_{\rm diff}\propto \elledd^{0.28}\mbh^{-1.04}Z^{-0.24}$, which has important observational
implications, indicating that the diffusion of hot gas is more sensitive to the SMBH mass than metallicity
and Eddington ratios. In light of $Z\propto L\propto \mbh\elledd$ from observations (Hamann \& Ferland
1999), we have $\dot{M}_{\rm diff}\propto \elledd^{0.04}\mbh^{-1.28}$. It is therefore expected that
the diffusion of the hot gas is not sensitive to the Eddington ratio of the accreting SMBH,
but $is$ moderately sensitive to the SMBH mass. For an individual quasar, $\dot{M}_{\rm inj}\propto Z^{-0.24}$,
implies that quasars in the late phase of a single episode should have relatively weaker outflows,
but higher metallicity. Last, the diffusion rates of the hot gas may be greatly enhanced if
$f_{\xi} = \xi f_*\gtrsim 10^{-3}\sunm^{-1}$ for a more top-heavy SF disk. It should be kept in mind
that equation (32) is a conservative estimate of the diffused hot gas from the disk.

We note that the hot gas inside the SF disk may have soft X-ray emission. Total emission from
the hot gas can be simply estimated by the free-free emission in the disk. We have
$L_{\rm HG}=\int 2\pi j(n_e,T_e)RHdR\approx 2.0\times 10^{40}\int r_3^{-1.36} dr_3\approx 5.0\times 10^{40}$
erg/s, where the emissivity of the free-free emission is
$j\approx 1.4\times 10^{-27}n_e^2T_e^{1/2}{\rm erg~s^{-1}~cm^{-3}}$.
The free-free emission in soft X-rays and radio would be negligible. It should be noted that in our model
the thermal diffusion is driven by a different mechanism than in the model
of winds driven by starbursts (Chevalier \& Clegg 1985). The present model stresses the role of evaporation
of the molecular clouds through their heating. In the starburst model, bubbles are energized as winds by
the kinetic energy of the starburst in a physically much thinner medium than the SF disk.

Finally the influence of the diffusion of hot gas on the disk can be estimated from the ratio of the disk
accretion rate to the mass rate of the diffusion. We find that $\dot{M}_{\rm diff}\sim 1 \sunmyr$ is
significantly smaller than the inflow mass rates ($\dot{M}_{\rm inflow}\sim 10\sunmyr$) (see Figure 2 in
Wang et al. 2010). The influence of the diffusion of hot gas on the disk can be generally neglected
in the star forming regions, but it might be important in the region around the self-gravity radius, where
the diffusion mass rates are comparable to the accretion rates onto the SMBH. In this region, the complicated
interaction between the warm gas and the accretion flow onto SMBHs may lead to some observable features
in X-ray spectra. We leave this for a future investigation. On the other hand, cooling of the hot winds may
be balanced by the efficient Compton heating driven by the X-ray photons emitted from the hot corona above
a Shakura-Sunyaev accretion disk. The final fate of the outflowing hot gas is a very complicated question,
since this gas will be subject to a complex array of processes as it moves out into the host galaxy. This
will be the main topic of a follow-up paper, but in the next section we outline some observable effects on
the BLR gas.

\section{Observational consequences for the BLR}
The BLR gas is assumed to be supplied by diffusion of the hot gas. There can be a factor of ten or more
metallicity difference between the inner and the outer parts of the SF disk, leading to observable 
consequences. This gas is then photoionized by the radiation from the accretion disk.
We can expect that the outer regions of the BLR are fed by the outer part of the SF disk. Since this
gas is farthest from the continuum source, it will have a relatively low value of the ionization
parameter, defined by $U=L_{\rm ion}/4\pi R^2c\langle\epsilon\rangle n$, where $L_{\rm ion}$ is the ionizing 
luminosity, $\langle\epsilon\rangle$ is the averaged energy of ionizing photons and $n$ is the density and 
temperature of the ionized gas, respectively.  Emission lines with low ionization
energy will be produced from this low ionization line (LIL) region.

The inner regions of the SF disk, being closer to the ionizing source, will have a higher value of
$U$, and thus produce high ionization lines (HILs). It should thus be possible to examine the
metallicity gradients through comparison of the LILs and HILs. Additionally, measuring metallicity
differences between the HIL and LIL regions can provide a rough estimate of the age of the single
episode of the SMBH activity, and potentially even reveal its evolutionary track. It is thus expected
that the metallicity measured from the LILs should be lower than the HILs. In principle, this theoretical
prediction can be examined by separate measurements of the metallicity in the HIL and LIL regions.

Generally, metallicity in AGNs is much higher than in their host galaxies. This is true even after
extrapolating the observed metallicity gradient in galaxies to the innermost regions. However, it is
much more difficult to measure the metallicity in AGNs than in stars. The original way to measure AGN
metallicity (Hamann \& Ferland 1993) was to use HIL ratios of \nvciv\, or \heii/\civ\ together with
detailed photoionization models. Later, broad-LIL ratios such as, \niiiciii\, and \niiioiii\,
were used together with the HIL lines to estimate the metallicity based on the LOC model (Hamann et al.
2000; Warner et al. 2003, but see the original suggestion of Shields 1976).
Warner et al. (2003, see Figure 10; and 2004) found that the metallicity
measured by \niiiciii\, is systematically lower than \nvciv\, by a factor of a few for a sample of
$\sim 800$ AGNs and quasars. Although they did not realize that these differences could be intrinsic
and instead took the averaged values to be the overall BLR metallicity, their measurements agree well
with the factor $\sim 10$ metallicity gradient across the SF disk that we find here. Assuming that both
kinds of metallicity indicators properly reflect the real metallicity, the systematic difference found
by Warner et al. (2003; 2004) clearly supports the idea that there is gradient of star formation within
the central 1pc region.

Additionally, the detailed nature of
the broad line regions is still very poorly understood.  However, the present simple model can
conveniently explain the significant difference of metallicity represented by \nvciv\, and
\niiiciii~ (Warner et al. 2003; 2004). Further high quality spectra of quasars are
needed to measure the metallicity accurately enough to determine the quasar's evolutionary status.

We would like to point out that there should be little or no mixing of metals between the LIL and HIL
regions, even though the dynamical time scales in the HIL and LIL are only a few tens of years. In a
new model of the BLR based on the present SF disk, we will show (Wang et al. 2011 in preparation) that
there is a mass circulation between the BLR and the SF disk, keeping a constant mass in the LIL regions,
whereas the cold clouds in the HIL could be blown away by the radiation pressure. The two regions maintain
a stationary state, showing different metallicity in LIL and HIL regions.

\section{Conclusions and Discussions}
In this paper, we have extended the study by Wang et al. (2010) of the properties of star forming
disks within the central 1 pc regions of quasars and AGNs, where star formation can drive an adequate
fuel supply to the SMBH through SNexp. We have made progress in the following areas:

\begin{itemize}
\item A metallicity equation is derived, unveiling a metallicity gradient in the star forming disk.
We find that metallicity can reach up to $\sim 10-20Z_{\odot}$ in the innermost regions of the disk
and a few times solar abundance in the outer part. As an integrated physical parameter, metallicity
is evolving even in disks that dynamically are in a steady state, and the metallicity can reach
saturation in a few $10^7$years. Metallicity can be used as a "clock" for indicating the age of a star
forming disk during one episode of SMBH activity.

\item Thermal diffusion with a mass rate of $\sim 1\sunmyr$ inevitably developes from the star
forming regions.
This rate is comparable to the outflow rates estimated from X-ray spectroscopic observations.
The hot gas diffused above the disk is likely to be the source of the observed broad-line-region gas.

\item The BLR should then share the metallicity gradient of the SF disk.
The metallicity gradient might already have been measured by the systematic difference in the
metallicity measured by the low (e.g. \niiiciii) and high (\nvciv) ionzation lines in AGNs and
quasars.
\end{itemize}

The properties of the star forming disk have important theoretical and observational consequences.
It is well known that AGNs have many separate ingredients, including the BLR (with separate
HIL and LIL regions), the NLR, the accretion disk, the self-gravitating disk and the torus. However,
the origins of these different components as well as the physical connections among them are extremely
poorly understood. Since the diffusion rates of the hot gas evaporated from the SF disk
are comparable with the accretion rates of the SMBHs, the presence of this gas will have significant
consequences. The ionized gas exposed to the AGN
irradiation will have a very complicated fate in light of AGN heating, cooling and coupling with its dynamics.
However, we speculate that the accumulation of the diffused gas continually supplied from the SF
disk can drive the formation of BLRs through dynamical and thermal instabilities. Dependence of the
properties of the diffused hot gas on the radius would then lead to onion-like structures within the
broad line regions.

It is often suggested that broad line regions, broad absorption lines, and
``warmer" absorbers are the result of outflows (e.g. Murray \& Chiang 1997). The present scenario provides
a gas supply to the broad
line regions, warmer absorbers etc. This points towards a unified model in which broad line
clouds, outflowing clouds and warmer absorbers are intrinsically linked with the star forming disk, and
hence with the accretion disk around the SMBHs. Such a model will be presented in a forthcoming paper
(Wang et al. 2011 in preparation).

A direct proof of the presence of massive stars in the core of a SF disk would be the
detection of the spectral features of red supergiants. In the near future, ALMA (Atacama
Large Millimeter Array) spectra of galactic nuclei should have the sensitivity to detect
the CO bandhead in the near IR that is predicted by our scenario.
Another approach would be to use powerful instruments to search for Ca {\sc ii} triplet
($\lambda\lambda 8498, 8542, 8662$\AA) absorption from the nuclear regions.

Finally, newly born AGNs and quasars may have rapidly growing accretion rates with self-similar
behaviors such as $\siggas\propto R^{\gamma_1}t^{\gamma_2}$ (Yu et al. 2005, but see Schawinski
et al. 2010). This would allow us to obtain the general solution for the metallicity evolution during
the early stage of the AGN phase. Interestingly, the metal-poor quasars have metallicities similar
to their host galaxies ($Z\le 0.6Z_\odot$), but they are only a small fraction of quasars (Fu \&
Stockton 2007). As has been suggested by Silk \& Rees (1998) and others, Ultra-Luminous Infrared
Galaxies (ULIRGs) with higher star formation rates (e.g. Laag et al. 2010) and low metallicity
are likely to be in the early stage of AGN activity, soon after their birth. Our work
here helps move us into a position to apply much more quantitative tests to this general idea.

\acknowledgements{We are grateful to the anonymous referee for a very useful report improving the
manuscript. H. Netzer and A. Laor are thanked for helpful suggestions and comments. We appreciate
the stimulating discussions among the members of IHEP AGN group. Conversations with J.-C. Wang are
appreciated. The research is supported by NSFC-10733010 and 10821061. JAB is grateful to
J.-M. Wang and the Institute of High Energy Physics of the Chinese Academy of Sciences for
their very gracious hospitality, and to NASA grant NNX10AD05G for financial support.}

\clearpage

\end{document}